\documentclass[aps,prl,twocolumn,groupedaddress]{revtex4-1}

\usepackage{graphicx}
\usepackage{amsmath}
\usepackage{amssymb}
\usepackage{siunitx}
\usepackage{array}
\usepackage{color}

\begin{document}

\title{High pressure effects on the  intermetallic superconductor Ti$_{0.85}$Pd$_{0.15}$}

\author{C.  Reyes-Dami\'an, F. Morales, Esmeralda Mart\'inez-Pi\~neiro, and Roberto Escudero}

\address{Instituto de Investigaciones en Materiales, Universidad Nacional Aut\'onoma de M\'exico, Ciudad de M\'exico, 04510 M\'exico }

\date{}

\begin{abstract}
This work reports superconductivity studies in the intermetallic  Ti$_{0.85}$Pd$_{0.15}$ performed in normal conditions and under hydrostatic pressure. The crystal structure of the compound has  a body centered cubic at room temperature and atmospheric pressure as unstable $\beta$-Ti phase. X-Ray diffraction pattern shows  space group I$m\bar{3}m$ with  parameter $a=3.2226(4)$ \r{A} and density around 5.6242 g/cm$^{3}$. The superconducting  transition temperature, T$_{C}=3.7$ K was determined from resistance, magnetization, and specific heat  measurements. The two critical magnetic fields, the coherence length, Ginzburg-Landau parameter, London penetration depth, the superconducting energy gap, the Debye temperature, the electron-phonon coupling constant and density of states at the Fermi level  were calculated. These parameters were obtained at ambient pressure. Under hydrostatic  pressure, the magnetic susceptibility measurements  show a small increment on T$_{C}$, the maximum T$_{C}=3.83$ K was obtained at the maximum applied pressure of  1.03 GPa. The slope calculated for T$_{C}$ as a linear function of pressure was about 0.14 K/GPa, possibly associated with an increase in the electronic density of states.
\end{abstract}

\keywords{Ti$_{0.85}$Pd$_{0.15}$ \and Superconductivity \and High Pressure \and Specific Heat }

\maketitle

\section{Introduction}
\label{intro}

Palladium and titanium are part of many superconductor compounds at low temperatures (10 K $<$). Some compounds are binary as Ti-Pd, Ti-Mo, Zr-Pd \cite{Poon}, Ti-FM (FM=Cr, Mn, Fe, Co and Ni) \cite{Matthias}, Pd-Te \cite{Raub,Tiwari,Amit} and Bi-Pd \cite{Jha}. Some materials has been studied with high pressure experiments, for example, external pressure on $\beta$-PdBi$_{2}$ supresses the superconducting transition \cite{Zhao,Pristas}.

In this research, we were focused in Ti-Pd system. The superconducting studies in these alloys were started 37 years ago. The studies at that time  showed an incomplete characterization of the material. Many compositions were attempted, some of the initial  compounds were carried out compositions as Ti$_{1-x}$Pd$_{x}$ in thin foils with  concentrations in the range 7-26\% atomic percent of Pd. These samples studied presented transition temperatures between 1.66 K for Ti$_{0.74}$Pd$_{0.26}$ and 3.7 K for Ti$_{0.85}$Pd$_{0.15}$, all samples had body-centered-cubic (bcc) structure, as $\beta$-Ti phase, with lattice parameter \textit{a} around 3.2 \r{A} \cite{Poon}. Due to the maximum T$_{C}$ for  Ti$_{0.85}$Pd$_{0.15}$ other studies were made \cite{Wong}. This alloy was synthesized using a modified piston-anvil technique. In this work was reported a transition temperature T$_{C}=3.82$ K  higher than 3.7 K \cite{Poon}. The upper critical field was determined using the Maki parameter $\alpha$ \cite{Maki64,Maki66} and the spin-orbit scattering parameter $\lambda_{s.o.}$  of the Werthamer-Helfald-Hohenberg (WHH) theory  \cite{WHH}. The Debye temperature ($\theta_{D}$) was approximated with the value for Ti$_{0.85}$Mo$_{0.15}$ as 295 K, and  electron-phonon coupling constant $\lambda_{e-ph}$=0.67 and densities of states at the Fermi level N(E$_{F}$)=0.53 $\frac{atates}{eV atom}$ were calculated.

Other studies in Ti-Pd alloys were performed in China in the same year, but in different compositions and in bulk samples. These samples were synthesized by arc melting. Only the samples with concentrations 0-29 atomic percent of Pd  presented superconductivity \cite{Luo}. Specific heat studies  performed in  Ti$_{0.8}$Pd$_{0.2}$  and Ti$_{0.92}$Pd$_{0.08}$ showed transition tempertures of 3.67 and 3.65 K respectively, values for $\theta_{D}$, electronic and phononic heat capacity coefficients ($\gamma$ and $\beta$), $\lambda_{e-ph}$, and N(E$_{F}$) were reported and compared with pure titanium values\cite{Jin}. The difference on the critical temperatures was associated to higher N(E$_{F}$) for alloys than N(E$_{F}$) of pure Ti.

Ti$_{0.85}$Pd$_{0.15}$ is the Ti-Pd alloy that shows the higher T$_{C}$ \cite{Poon} among these alloys, however, its superconducting characteristics are studied scarcely.
In this work we present a more complete characterization of this superconducting alloy. The studies include electric measurements, magnetic measurements, specific heat measurements and high pressure measurements. From magnetic and electric measurements we determined the magnetic critical fields, H$_{C1}$ and H$_{C2}$, and Ginzburg-Landau parameters. From the specific heat measurements we determined the superconducting energy gap $2\Delta$, and normal state parameters as $\theta_{D}$, $\gamma$, $\beta$, and the electronic density of states at the Fermi level.
The high pressure experiments show that T$_{C}$ increases as the pressure increases in the pressure range studied.

\section{Experimental Details}
\label{Exp}

The analyzed sample with stoichiometry  Ti$_{0.85}$Pd$_{0.15}$ was prepared in  arc  furnace under a high purity argon atmosphere. The starting materials were Pd powders (Sigma Aldrich, 99.999\%) and Ti pellets (ESPI, 99.97\%). The Pd powder was pelletized, whereas the  Ti pellets were cut  in small pieces. Both, powders and pieces, were introduced into  the furnace chamber with an argon atmosphere. The synthesis was performed in several stages and annealed several times, at least three times, in order to obtain optimum  homogenized samples. The final samples presents a spherical shape. These samples were cut with a diamond wheel for easy handling, and powder of the composition was collected.

The crystal structure was determined  and analyzed from the X-ray diffraction pattern obtained by the powder method,  performed at room temperature in a D5000-Siemens diffractometer with the Bragg-Brentano geometry and CoK$_{\alpha}$ radiation ($\lambda=1.7903$ \r{A}). Rietveld refinement was made with GSAS-II software \cite{gsas}.

The superconducting characteristics were determined from magnetic, resistance, and heat capacity measurements at low temperatures.
Magnetic susceptibility of samples  ($\chi$(T)) was measured with a magnetic field, H=10 Oe, using  the Zero Field Cooling (ZFC) and Field Cooling (FC) protocols. The lower  Critical magnetic field H$_{C1}$ was obtained  from isothermal magnetization measurements as a function of the  magnetic field. We used a Quantum Design magnetometer (MPMS) for these studies.

The  electrical resistance as a function of temperature (R(T)) was measured using the four wires configuration in  a Quantum Design Physical Properties Measurement System (PPMS) at zero magnetic field  from room temperature to 2 K. In order to determine the upper critical field H$_{C2}$, R(T) was  measured close to T$_{C}$ at different magnetic fields.

The  specific heat  behavior gives information about the bulk superconductivity. In the superconducting state it is possible to obtain microscopic characteristics, as the size of the energy gap (2$\Delta$),  $\lambda_{e-ph}$ and N(E$_{F}$). The specific heat was measured from 30 to 2 K using the thermal relaxation method in the PPMS.

Lastly, $\chi$(T) measurements at high pressure were performed in the MPMS using  a  Quantum Design CuBe high pressure cell (HPC) with H=10 Oe.In adittion to the sample a piece of lead was included, as a manometer, into the pressure cell \cite{Clark}. The $\chi$(T) analysis for Ti$_{0.85}$Pd$_{0.15}$ at high pressure was made subtracting the lead contribution from the values of susceptibility. The Daphne 7373 oil  was the pressure transmitting medium because it has not a magnetic contribution \cite{Esmeralda}.

\section{Results and Discussion}
\label{Res}

\subsection{X-Ray Diffraction and Structural Characterization}
\label{X-Ray}

\begin{figure}[h!]
\includegraphics [width=0.5\textwidth]{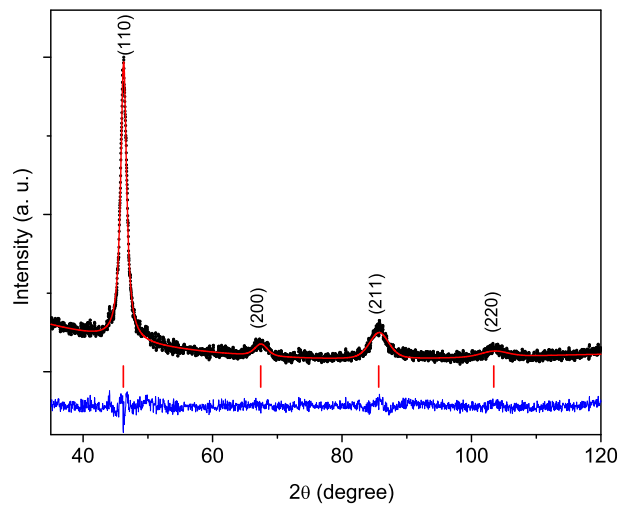}
\caption{(Color online) X-ray diffraction pattern of Ti$_{0.85}$Pd$_{0.15}$ (points) measured with CoK$_{\alpha}$ radiation, and Rietveld refinement calculated (solid line). Vertical lines are the reflection positions. And the bottom line is the difference between experimental and calculated patterns}
\label{Fig:1}
\end{figure}

Figure \ref{Fig:1} shows the experimental X-Ray pattern, and calculated pattern by the Rietveld refinement method. The cell parameter, crystal structure, and atomic positions were taken from the database PDF4 (ICDD) with the entry (01-072-2907), that correspond to Ti$_{0.8}$Pd$_{0.2}$ in order to perform the refinement. The Rietveld refinement gave the following parameters; $GOF=1.207$ and $R_{WP}=12.89\%$, that guarantee the accuracy of the calculus. The crystal structure of Ti$_{0.85}$Pd$_{0.15}$  has cubic symmetry with spatial group I$m\bar{3}m$, and  lattice parameter  $a=3.2226(4)$ \r{A}, this value is in agreement with previous report \cite{Poon}. It is noteworthy that the lattice parameter, in the Ti-Pd alloys, shows an increment as the Pd increases until $a$ reaches a maximum at 15 at. \% Pd, after that, $a$ decreases \cite{Poon}.
The calculated mass density was 5.6242 g/cm$^3$.

\subsection{Superconducting State}
\label{Super}

\begin{figure}[h!]
\includegraphics[width=0.5\textwidth]{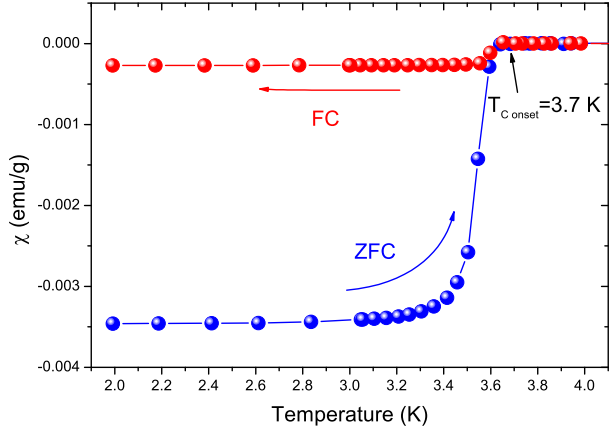}
\caption{(Color online) $\chi(T)$ for the superconducting composition  Ti$_{0.85}$Pd$_{0.15}$ measured at 10 Oe in ZFC and FC  modes. The arrow marks the onset temperature at 3.7 K}
\label{Fig:2}
\end{figure}

T$_{C}$ of Ti$_{0.85}$Pd$_{0.15}$ was determined  from magnetic susceptibility measures as a function of temperature. $\chi$(T) was measured in ZFC and FC modes using 10 Oe as applied magnetic field. The $\chi$(T) behavior at low temperatures is shown in Fig. \ref{Fig:2}. There the onset temperature of the superconducting state is observed at 3.7 K.

Also the superconducting transition temperature can be determined from electrical resistance as a function of temperature R(T), T$_{C}$ was defined as the onset of the transition at T$_{C_{onset}}$=3.7 K. We defined T $_{C_{zero}}$=3.5 when R(T) is zero. The difference between these two critical temperatures is the superconducting transition width $\Delta$T=0.2 K.

\begin{figure}[h!]
\includegraphics[width=0.5\textwidth]{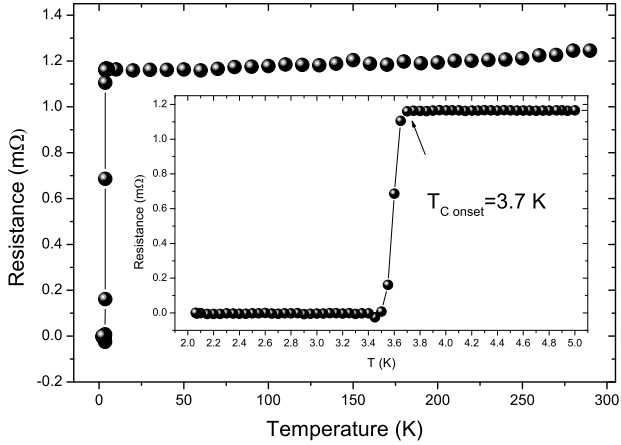}
\caption{Resistance behavior measured without applied magnetic field. The inset shows  the behavior at low temperatures, and the arrow marks the onset transition temperature at 3.7 K}
\label{Fig:3}
\end{figure}

\begin{figure}[h!]
\includegraphics[width=0.5\textwidth]{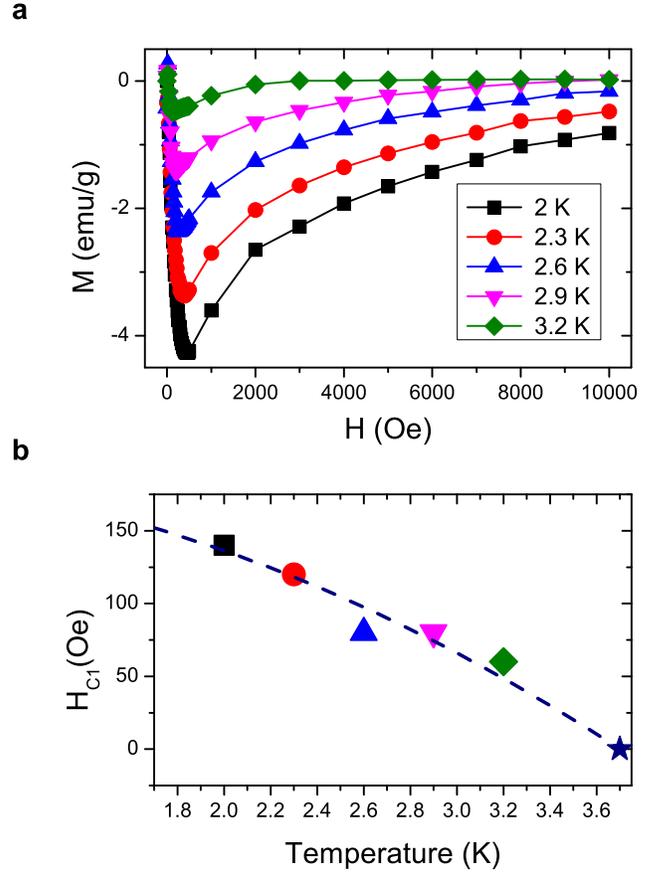}
\caption{ \textbf{a}(Color online) Magnetization curves as a function of magnetic field (H) at different temperatures, as indicated. \textbf{b} H$_{C1}$ was determined at constant temperature as the field where the diamagnetic curve starts to deviate from linearity for each isothermal.  H$_{C1}$(0) was obtained from the parabolic fit (dashed line)}
\label{Fig:4}
\end{figure}

With isothermal measurements of magnetic susceptibility as a function of magnetic field, we determined the lower critical field values. The magnetic field H was increased from 0 up to 10,000 Oe, as seen in Fig. \ref{Fig:4}a. In order to determine H$_{C1}$(T), we used the points where the diamagnetic curve starts to deviate from linearity, for each isothermal. The  H$_{C1}$ behavior is shown in the figure \ref{Fig:4}b, the data was fitted  to a parabolic  equation;  H$_{C1}$(T)=H$_{C1}$(0)[1-(T/T$_{C}$)$^{2}$]. From this fitting we determined  H$_{C1}$(0)=193 Oe.

\begin{figure}[h!]
\includegraphics[width=0.5\textwidth]{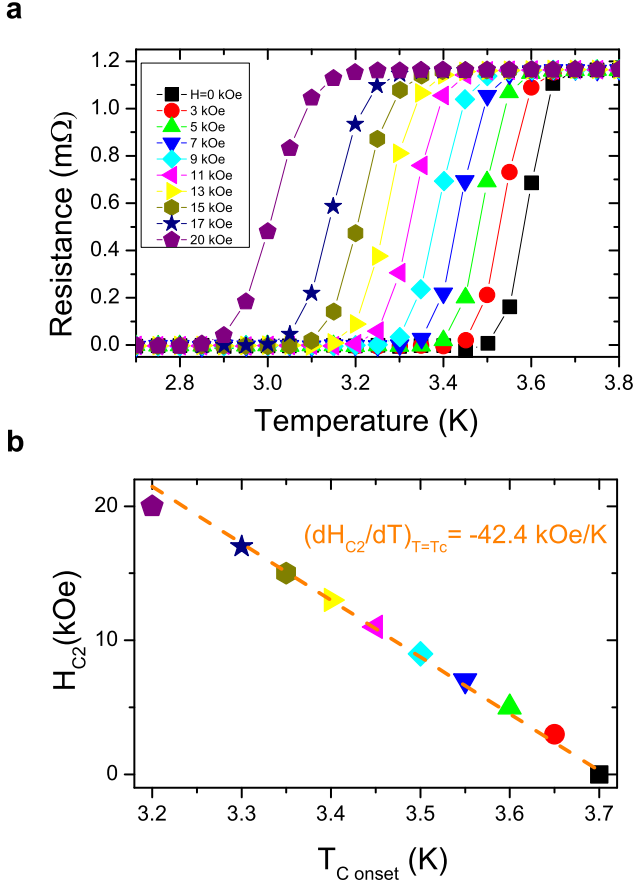}
\caption{\textbf{a} (Color online) Electrical Resistance as a function of temperature at different applied magnetic fields H. \textbf{b}  The transition temperature was taken at the onset  for each curve. The upper field,  H$_{C2}$(0)  was calculated with the WWH approximation using the slope from the linear fit (dashed line)}
\label{Fig:5}
\end{figure}

The upper critical fields H$_{C2}$(T) values were determined from the R(T) curves measured at different applied magnetic field, see Fig. \ref{Fig:5}a, the applied magnetic field is set as H$_{C2}$ for a given T$_{C onset}$. The relation between T$_{C onset}$ and H$_{C2}$ is practically linear, as it can be seen in  Fig. \ref{Fig:5}b . H$_{C2}$(0) was calculated using the approximation of Werthamer-Helfald-Hohenberg (WHH)\cite{WHH}:
\begin{equation*}
\textrm{H}_{C2}(0)=-0.693\textrm{T}_{C}\left(\frac{\textrm{dH}_{C2}}{\textrm{dT}}\right)_{\textrm{T=T}_{C}},
\end{equation*}
where $(\frac{\textrm{dH}_{C2}}{\textrm{dT}})_{\textrm{T=T}_{C}}$ represents the slope of the linear fit of data close to  T$_{C}$. The calculated value for H$_{C2}$(0) is 108.7 kOe.

With the values of  H$_{C1}$(0) and H$_{C2}$(0)we calclated the Ginzburg-Landau parameters: the  coherence length $\xi_{GL}$ using  equation $\xi_{GL}=\sqrt{\Phi_{0}/2\pi \textrm{H}_{C2}}$ where $\Phi_{0}$ is the quantum fluxoid , the G-L parameter $\kappa$ determined with $\textrm{H}_{C1}/\textrm{H}_{C2}\approx ln(\kappa)/2\sqrt{2}\kappa^{2}$, and  the London penetration depth $\lambda_{L}$ from the relation  $\kappa=\lambda_{L}/\xi_{GL}$ \cite{Esmeralda}. The values are in the Table \ref{tab:1}

\begin{figure}[h!]
\includegraphics[width=0.5\textwidth]{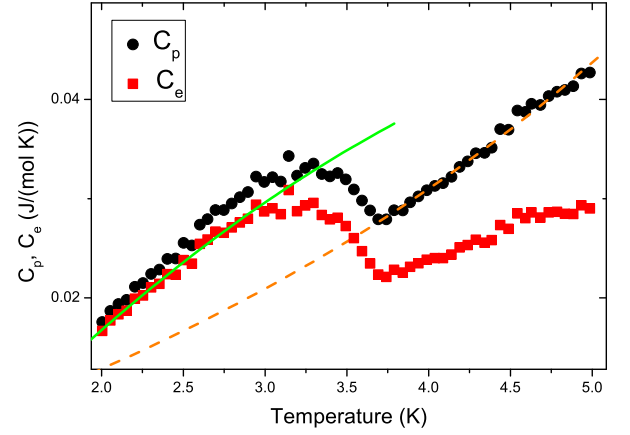}
\caption{(Color Online) C$_{p}$ and the electronic contribution  C$_{e}$ as a function of temperature at H=0 Oe were plotted. C$_{p}$ was fitted above the transition temperature using  the equation C$_{P}$= $\gamma$ T+$\beta$ T${^3}$ in order to obtain $\gamma$ and $\beta$ values (dashed line). The energy gap can be calculate using the equation $\textrm{C}_{e}=Ae^{(-\Delta/k_{B}\textrm{T}})$ from C$_{e}$ (solid line)}
\label{Fig:6}
\end{figure}

Figure \ref{Fig:6} shows specific heat (C$_{P}$) and electronic contribution to C$_{P}$ (C$_{e}$) for Ti$_{0.85}$Pd$_{0.15}$ at low temperatures. In order to separate the different contribution to C$_{P}$ we use the equation, C$_{P}$(T)= $\gamma$ T+$\beta$ T${^3}$, where $\gamma$T is C$_{e}$ and $\gamma$ is the Sommerfeld constant. The second term $\beta$T${^3}$ is the low temperature Debye approximation \cite{Tari}. We fitted C$_{P}$ data above T$_{C}$ ( dashed line) and obtained values for $ \gamma =5.98$ mJ/(mol K$^{2}$) and $\beta=0.11$ mJ/(mol K$^{4}$). The electronic contribution can be determined subtracting the Debye term from the experimental data. With C$_{e}$ is possible knows the size of the energy gap, fitting C$_{e}$ data  with $\textrm{C}_{e}=Ae^{(-\Delta/k_{B}\textrm{T})}$ below the maximum of the peak (solid line). The calculated $\Delta$ value was 0.295 meV, it gives 2$\Delta$/k$_{B}$T=1.85 lower than BCS prediction of 3.52 \cite{BCS}. This ratio indicates that Ti$_{0.85}$Pd$_{0.15}$ is a superconducting material in the weak coupling electron-phonon regime.

The Debye temperature ($\theta_{D}$)  was calculated from the relation $\beta=12 \pi^{4}R/5\theta_{D}^{3}$, where \textit{R} is the ideal gas constant, the calculated value of $\theta_{D}$ was 260.5 K. This value is lower than 295 K reported previously \cite{Wong}, however, in this they assume this value has similar value as in the Ti$_{0.85}$Mo$_{0.15}$ alloy. With this consideration, we can not make a fair comparison with the values of $\lambda_{e-ph}$ and  N(E$_{F}$) reported for Ti$_{0.85}$Pd$_{0.15}$.

The electron-phonon coupling constant ($\lambda_{e-ph}$) can be calculated using the empirical equation proposed by McMillan \cite{McMillan},

\begin{equation*}
\lambda_{e-ph} =\frac{1.04+ \mu ^{*} ln(\theta _{D}/1.45\textrm{T}_{C})}{(1-0.62 \mu ^{*})ln(\theta _{D}/1.45\textrm{T}_{C})-1.04},
\end{equation*}

were $\mu ^{*}$ is the Coulomb pseudopotential. Using this equation and the values of $\theta_D$ and T$_{C}$ obtained for Ti$_{0.85}$Pd$_{0.15}$, presented in Table \ref{tab:1}, and  $\mu^{*}=0.1$ it was obtained $\lambda_{e-ph} = 0.55$, similar to 0.5 reported previously \cite{Jin}. This value indicates a weak coupling electron-phonon interaction.
Lastly, we calculated the N(E$_{F}$) value from the relation with $\gamma$:

 N(E$_{F}$)=$3\gamma/2\pi^2k_{B}^2(1+\lambda_{e-ph})$.

We summarize the determined superconducting and normal parameters  in Table \ref{tab:1}.

\begin{table*}[t!]
\begin{center}
\caption{Parameters determined, at zero pressure, for the intermetallic alloy Ti$_{85}$Pd$_{15}$ in the superconducting and normal state.}
\label{tab:1}
\begin{tabular}{cccccccccccccc}
\hline\noalign{\smallskip}
Composition&T$_{c}$ &$H_{c1}$(0) & $H_{c2}$(0)&$\xi_{GL}$ & $\kappa$ &$\lambda_{L}$ & $\Delta$ & $2\Delta/K_BT_C$ & $\gamma$ & $\beta$ & $\theta_D$& $\lambda_{e-ph}$& N(E$_{F}$)\\
& K & kOe & kOe & nm &  & nm  & meV &  & $\frac{mJ}{molK^{2}}$ & $\frac{mJ}{molK^{4}}$ & K & & $\frac{states}{eV atom}$ \\
\noalign{\smallskip}\hline\noalign{\smallskip}
Ti$_{0.85}$Pd$_{0.15}$&3.7&0.193 & 108.7 &5.5& 25.38 & 140 &  0.295 & 1.85 & 5.98 & 0.11 & 260.5&0.55&0.82 \\
\noalign{\smallskip}\hline
\end{tabular}
\end{center}
\end{table*}

The value of $\kappa$ indicates that Ti$_{0.85}$Pd$_{0.15}$ is a type-II superconductor{\color{red},} and the parameters $\lambda_{L}$ and $\xi_{GL}$ are in typical values \cite{Fickket}. The upper critical field was 1.6 times the value reported for Ti$_{0.85}$Pd$_{0.15}$ H$_{C2}(0)\approx67$ kOe \cite{Wong}. The ratio 2$\Delta$/k$_{B}$T$_{C}$=1.85 value is related with a superconducting material in the weak coupling electron-phonon limit, $\lambda_{e-ph}$=0.55 confirms the weak coupling.

\subsection{Superconducting State at High Pressure}
\label{High P}

In the previous section, we reported the superconducting characteristics of Ti$_{0.85}$Pd$_{0.15}$ determined by $\chi$(T), R(T) and C$_{p}$(T)  at atmospheric pressure. Now the hydrostatic pressure effect on T$_{C}$ can  be  studied. For this purpose   the sample (6.2 mg) and 1 mm of Pb wire (3.1 mg) were introduced in the HPC. From the mgnetization measurement, without applied pressure, the T$_{C}$ of the sample and Pb were determined as 3.7 K and 7.19 K, respectively, using H=10 Oe and a small temperature step size ($\textrm{dT}\approx0.02$ K) in ZFC mode. The HPC was compressed, and the pressure on the sample region rise, to determine the pressure value, we use the know change rate {$\textrm{dT}_{C}/\textrm{dP}=-0.379$ K/GPa for Pb \cite{Clark}.
Fig. \ref{Fig:7} shows $\chi$(T) of Ti$_{0.85}$Pd$_{0.15}$ at four different pressure and the reference at P=0 GPa. The inset of Fig. \ref{Fig:7} shows an amplification of the data around the transition temperature, there it is evident the pressure effects on T$_{C}$.

\begin{figure}[h!]
\includegraphics[width=0.5\textwidth]{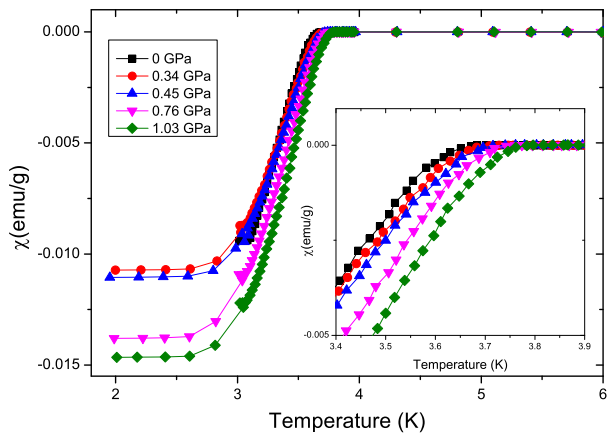}
\caption{ (Color online)Ti$_{0.85}$Pd$_{0.15}$ ZFC measurements at high pressures, as indicated. The inset shows the behavior close the transition temperature}
\label{Fig:7}
\end{figure}

In Fig. \ref{Fig:8} is plotted the critical temperature as a function of pressure of Ti$_{0.85}$Pd$_{0.15}$. The dashed line is a linear fit of the data, the slope of this line is $\textrm{dT}_{C}/\textrm{dP}=0.14$ K/GPa.

 \begin{figure}[h!]
\includegraphics[width=0.5\textwidth]{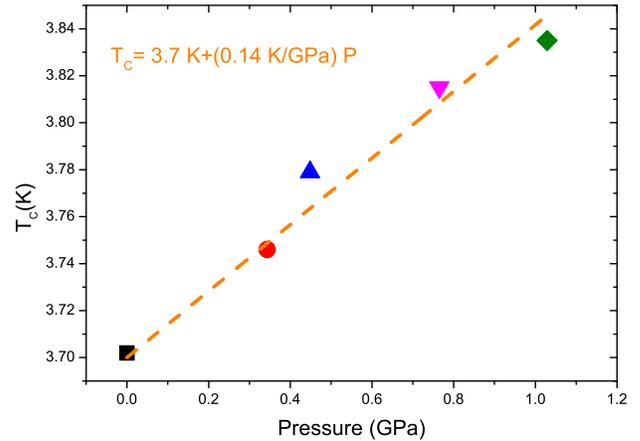}
\caption{(Color Online)T$_{C}$ as a function of pressure was plotted. Dashed line is the linear fit, were the slope is the change rate of T$_{C}$ with pressure}
\label{Fig:8}
\end{figure}

Due to the high concentration of palladium in Pd-Ti alloys, we compared  the beavior of T$_{C}$ at high pressures for Ti$_{0.85}$Pd$_{0.15}$ with pure Ti phases.
The direct comparison could be with $\beta$-Ti phase because it has the same bcc crystal structure, but has no studies at high pressure. This phase is unstable and the calculus of his properties at atmospheric pressure are made on Ti$_{1-x}$Mo$_{x}$ alloys and extrapolating to zero Mo content \cite{Baconyi,Collings}.
The $\alpha$-Ti phase has a hexagonal close-packed structure and  positive $\textrm{dT}_{C}/\textrm{dP}=0.55$ K/GPa, but in a pressure region of 1.5-2.5 GPa, the ratio below 1.5 GPa was almost zero \cite{Brandt}.
Another phase with hexagonal structure, $\omega$-Ti  has a change rate $\textrm{dT}_{C}/\textrm{dP}=0.07$ K/GPa at pressures between 40-60 GPa, with a maximum  T$_{C}$ of 3.35 K at 60 GPa \cite{Bashkin}.
Note that these Ti phases show an important increment of T$_{C}$ at high pressures. This behavior has been interpreted as result of a continuum transference of $s$ electrons to the $d$ electronic band with applied pressure, particularly in $\alpha$-Ti with a rate 0.002 electros/atom per GPa  \cite{Bashkin,Gyanch}. The same electronic mechanism has been proposed for vanadium at high pressure. Due to the pressure-induced $s$-band to $d$-band transfer, Ti become similar to the nearby element V \cite{Hamlin}. It is reasonable that the T$_{C}$ increment with pressure in Ti$_{0.85}$Pd$_{0.15}$ could be the same mechanism.

\section{Conclusions}
\label{Con}

This work presents the characteristics of the superconductor Ti$_{0.85}$Pd$_{0.15}$.  The sample is a single phase determined by X-Ray diffraction and Rietveld refinements, the  crystal structure is bcc with lattice parameter $a=3.2226(4)$ \r{A} and space group I$m\bar{3}m$. The Ti$_{0.85}$Pd$_{0.15}$ alloy has T$_{C}=3.7$ K and is a type-II superconductor  with H$_{C1}=193$ Oe and H$_{C2}=108.7$ kOe. The bulk superconductivity  was determined by heat capacity measurement. From heat capacity as a function of temperature were calculated microscopic characteristic as the superconducting energy gap, $\lambda_{e-ph}$ and N(E$_{F}$). The values of 2$\Delta$/k$_{B}$T=1.85 and $\lambda_{e-ph}$=0.55 indicates that Ti$_{0.85}$Pd$_{0.15}$ is a superconducting material in the weak coupling electron-phonon regime. From $\chi$(T) at high pressure, we observed that T$_{C}$ increases with applied pressure, the change of T$_{C}$ with pressure is $(\frac{\textrm{dT}_{C}}{\textrm{dP}})=0.14$ K/GPa, between  0 and 1 GPa. This increment on T$_{C}$ could be due to an increment of N(E$_{F}$) with the pressure.

\begin{acknowledgements}
We thanks to A. Bobadilla for helium provisions, to A. Pompa and A. Lopez for helps in software processes. Also to DGAPA-UNAM IT100217, and CONACyT for the scholarship to Carlos Reyes-Damián.
\end{acknowledgements}

\end{document}